# Topological Semimetal Transport Modulated by Interstitial Fe in Ba(Fe$_{1-x}$Co$_x$)$_{2+\delta}$As$_2$ Superconductors

Ze-Xian Deng[1], Qiang-Jun Cheng[1], Jing Jiang[2,3], Yong-Wei Wang[1], Xi Zhou[1], Ming-Qiang Ren[1,4], Cong Cong Lou[1], Xiao-Xiang Chen[1], Bin-Jie Wu[1], Zeng-Wei Zhu[5], Qing-Hua Zhang[6], Lin Gu[7], Ding Zhang[1,8,9], Kai Liu[2,3], Xu-Cun Ma[1,8], Qi-Kun Xue[1,4,8,9,10], and Can-Li Song[1,8]

[1]*Department of Physics and State Key Laboratory of Low-Dimensional Quantum Physics, Tsinghua University, Beijing 100084, China*

[2]*School of Physics and Beijing Key Laboratory of Opto-electronic Functional Materials & Micro-nano Devices, Renmin University of China, Beijing 100872, China*

[3]*Key Laboratory of Quantum State Construction and Manipulation (Ministry of Education), Renmin University of China, Beijing 100872, China*

[4]*Department of Physics, Southern University of Science and Technology, Shenzhen 518055, China*

[5]*Wuhan National High Magnetic Field Center and School of Physics, Huazhong University of Science and Technology, Wuhan 430074, China*

[6]*Institute of Physics, Chinese Academy of Sciences, Beijing 100190, China*

[7]*State Key Laboratory of New Ceramics and Fine Processing, School of Materials Science and Engineering, Tsinghua University, Beijing 100084, China*

[8]*Frontier Science Center for Quantum Information, Beijing 100084, China*

[9]*Beijing Academy of Quantum Information Sciences, Beijing 100193, China*

[10]*Hefei National Laboratory, Hefei 230088, China*

Topological semimetals are renowned for exhibiting large, unsaturated magnetoresistance arising from ultrahigh carrier mobility and electron-hole compensation. However, such behaviors remain poorly understood in iron-based superconductors that have been recently recognized to harbor rich nontrivial topology. Here, we combine angle-resolved magneto-transport measurements with first principles calculations to reveal the emergence and tunability of topological semimetals in ferropnictide Ba(Fe$_{1-x}$Co$_x$)$_{2+\delta}$As$_2$ epitaxial films, modulated by interstitial Fe. These states exhibit ultralow residual resistivity, coexisting high-mobility electron and hole carriers, and linear positive magnetoresistance below 110 K. Remarkably, the magnetoresistance becomes more pronounced when the magnetic field is applied parallel to the film plane, reaching an unsaturated 1206% at 56 T. Furthermore, superconductivity persists in these ferropnictide films, establishing them as a tunable platform for investigating the interplay among electron correlation, topology, and superconductivity.

*Corresponding authors. Emails: clsong07@mail.tsinghua.edu.cn, kliu@ruc.edu.cn, xucunma@mail.tsinghua.edu.cn, qkxue@mail.tsinghua.edu.cn

Topological quantum states of matter, characterized by nontrivial band topology and resilience to local perturbations [1], were initially formulated within a single particle framework but are now increasingly recognized in correlated electron systems [2-4]. This paradigm shift has expanded the landscape of topological phases and stimulated efforts to identify correlation-driven topological phenomena [5]. A central focus of this pursuit is the emergence of nontrivial band topology in high-temperature ($T_c$) superconductors [6-9], where its coupling to unconventional pairing and electron correlations may generate novel quantum states. In this context, iron-based superconductors (IBSs) are particularly appealing due to their multiband electronic structure and sizable spin-orbit coupling, which host topological surface states (TSSs) associated with a bulk topological insulator (TI) phase [7-9], three-dimensional topological Dirac semimetals (TDSs) [9, 10] and higher-order topological phases [11]. Recent angle-resolved photoemission spectroscopy experiments have mapped the TSSs and TDSs in IBSs [10, 12], and Majorana zero modes (MZMs) associated with TSSs have been reported in magnetic vortices [10, 13-16].

Despite the advance, macroscopic quantum-transport signatures of topological states in IBSs remain elusive. Transverse magnetoresistance (MR) with the magnetic field $\mu_0 H$ perpendicular to the current ($I$) were sporadically reported in Fe(Se,Te) [9], BaFe$_2$As$_2$ single crystals and their doped derivatives [17-20], but both the magnitude and temperature dependence vary substantially from sample to sample, with their underlying origin largely unexplored. Moreover, longitudinal MR ($\mu_0 H // I$), a major hallmark for identifying topological transport [21], has remained elusive to date. Here we report unsaturated, linear positive MR at the normal state of interstitial-Fe (Fe$_i$)-modulated Ba(Fe$_{1-x}$Co$_x$)$_{2+\delta}$As$_2$ (BFCA) superconducting films for both longitudinal and transverse magnetic-field geometries. Together with the ultralow residual resistivity, the coexistence of high-mobility electron and hole carriers, and density functional theory (DFT) calculations, our results provide compelling evidence for TDS-driven topological transport in IBSs.

The BFCA films were grown on insulating SrTiO$_3$(001) substrates by molecular beam epitaxy (MBE), as described in detail elsewhere [22, 23] and in the Supplemental Material [24]. The crystal structure was analyzed by X-ray diffraction (XRD) via a Rigaku X-ray diffractometer with Cu $K_{\alpha 1}$ radiation and by an aberration-corrected scanning transmission electron microscopy (TEM, JEOL ARM200CF). Electrical resistivity and Hall effect measurements were performed using the standard four-probe technique in a standard physical property measurement system (PPMS) and at the Wuhan National High Magnetic Field Center. DFT calculations were carried out by using the VASP package [25, 26], with a plane-wave kinetic energy cutoff of 520 eV and an $11 \times 11 \times 11$ k-point mesh for Brillouin-zone sampling [24].

Interstitial Fe (more precisely, excess Fe and Co atoms) are incorporated into the As layers [Fig. 1(a)] by reducing the Ba flux relative to that used for the codeposition growth of $Ba(Fe_{1-x}Co_x)_2As_2$, while keeping the Fe and Co fluxes, as well as their ratio, fixed. To isolate and investigate the effects of $Fe_i$, the Co concentration $x$ was chosen near the optimal doping range, and the film thickness was kept at 10 unit cells (~ 13.2 nm). Figure 1(b) shows XRD spectra of BFCA films as a function of the nominal $Fe_i$ content $\delta$. The main diffraction peaks progressively shift towards low angles with increasing $\delta$, indicating a lattice expansion along the $c$ axis associated with the $Fe_i$ incorporation. High-resolution TEM measurements reveal that $Fe_i$ atoms are non-uniformly distributed at interstitial sites within the As planes [Figs. 1(c) and 1(d)], similar to a recent observation in $Fe_{1+y}Te_{1-x}Se_x$ [27]. This finding is further supported by statistical analysis of the intensities of the atomic columns corresponding to $Fe_i$ and planner Fe [Fig. 1(e)]. While the planer Fe columns exhibit a Gaussian-shaped intensity distribution, as anticipated, the $Fe_i$ columns display asymmetric, double-Gaussian distributions, reflecting the inhomogeneous spatial distribution of $Fe_i$.

Figure 2(a) displays the temperature dependence of the logarithmic resistivity $\rho_{xx}$, measured at zero magnetic field for two representative $Fe_i$-modulated BFCA films. At low temperatures, the $\rho_{xx}$ is strongly suppressed, giving rise to a markedly enhanced residual resistivity ratio (RRR) that correlates closely with the $Fe_i$ content [Figs. S1(a-c)]. For sample #1 ($\delta = 0.50$ and $x = 6.5\%$), the RRR, defined as the ratio of $\rho_{xx}(300\ K)$ to $\rho_{xx}$ at the superconducting onset temperature $T_{c,\text{onset}} \approx 19.4$ K, reaches a value as high as 776, with $\rho_{xx}(T_{c,\text{onset}})$ as low as 27 n$\Omega\cdot$cm. Moreover, $\rho_{xx}$ exhibits a clear $T^2$ temperature dependence at low temperatures [Fig. 2(b)], extrapolating to a $\rho_{xx}(0) \approx 1.2$ n$\Omega\cdot$cm. This corresponds to an extrapolated RRR on the order of $10^5$. Such an ultralow residual resistivity and extraordinarily large RRR were rarely observed even in high-purity stoichiometric compounds, and are therefore highly unexpected in nonstoichiometric BFCA films containing a substantial density of randomly distributed $Fe_i$ and Co dopants.

Figure 2(c) summarizes the conductivity evaluated at $T_{c,onset}$ (i.e. $1/\rho_{xx}(T_{c,onset})$), plotted against the RRR for the BFCA films studied, together with representative superconductors for comparison. While BFCA films without or with a small amount of Fe$_i$ ($\delta$ = 0.2 in Fig. S1(b)) exhibit conductivity and RRR comparable to those of bulk Ba(Fe$_{1-x}$Co$_x$)$_2$As$_2$ crystals [20], BFCA films with larger $\delta$ invariably display ultrahigh low-temperature conductivity (equivalently, ultralow residual resistivity) and exceptionally large RRR. Notably, the residual resistivity (RRR) in these BFCA films is significantly lower (higher) than that of all well-known superconductors, including high-quality stoichiometric compounds such as LiFeAs [28], CsV$_3$Sb$_5$ [29] and KFe$_2$As$_2$ [30], and is comparable to that of high-purity cooper at 20 K. Importantly, although superconductivity is moderately suppressed, it remains robust across all BFCA films studied [Fig. S1(a-c)]. These results, established with statistical significance, demonstrate that ultralow residual resistivity and large RRR are intrinsic properties of Fe$_i$-modulated BFCA superconductors.

Accompanying the ultralow residual resistivity, another prominent feature of the Fe$_i$-modulated BFCA films is the emergence of sizable magnetoresistance below 110 K, as indicated by red arrows in Fig. 3(a). Despite some sample and $x$-dependent variations, the magnetic field-induced resistivity enhancement at the normal state becomes more pronounced upon cooling and with increasing Fe$_i$ content [Figs. S1(d-f)]. Figure 3(b) exemplifies the magnetic-field ($\mu_0H$)-dependent transverse MR for sample #2 at various temperatures above $T_c$, where the MR($\mu_0H$), in %, is defined as MR = [$\rho_{xx}(\mu_0H) - \rho_{xx}(0)]/\rho_{xx}(0)\cdot 100\%$ and the field is applied along the crystallographic $c$ axis. At low $\mu_0H$, the MR exhibits a weak quadratic curvature, while at higher fields it crosses over into an unsaturated, linear positive dependence extending over the entire accessible field range. This crossover and the linear-in-field character are further highlighted by the field derivative $d$MR/$d\mu_0H$ [Fig. 3(c)], which approaches a constant value above a crossover field $\mu_0H^* \sim 1.5$ T, a hallmark of linear MR. These behaviors stand in sharp contrast to that of common semimetals [50], where the MR generally follows a quadratic dependence on $\mu_0H$ over the entire field range.

To further elucidate the origin of the linear MR and its angular dependence, we systematically measured the MR with the magnetic field applied along different crystallographic orientations, as illustrated by the schematic measurement geometry [inset of Fig. 3(b)]. Figure 3(d) shows the transverse MR measured at 30 K as the magnetic field is rotated from the crystallographic $c$ axis toward the in-plane $b$ axis. With increasing polar angle $\theta$, the MR is apparently enhanced, indicating its pronounced sensitivity to the field orientation relative to the film plane. When the magnetic field is fully aligned within the film plane, the transverse MR reaches its maxima and remains unsaturated up to the highest accessible pulsed field of 56 T [Figs. 3(e), S2 and S3], with a value of ∼ 1206% at 27 K. Strikingly, under in-plane magnetic fields, the MR measured in the longitudinal ($H//I$) and transverse ($H \perp I$) configurations exhibits comparable magnitudes, both exceeding the transverse MR obtained with out-of-plane magnetic fields. This observation is highly unconventional, as classical orbital MR induced by the Lorentz force is expected to be strongly suppressed in the longitudinal geometry, as observed in stoichiometric parent $BaFe_2As_2$ films [Fig. S4]. The comparable enhancement of longitudinal and transverse MR for in-plane magnetic fields therefore points to an anomalous, nonclassical transport mechanism in $Fe_i$-modulated BFCA films.

Hall effect measurements provide additional insight into the carrier dynamics responsible for the unusual transport behaviors in the $Fe_i$-modulated BFCA films. Figure 4(a) shows the magnetic-field dependence of the Hall resistivity $\rho_{xy}$ measured at various temperatures for the representative BFCA film with $\delta = 0.52$. At high temperatures, $\rho_{xy}$ exhibits a nearly linear dependence on $\mu_0 H$, similar to that observed in $Fe_i$-free BFCA films over the temperature range from room temperature down to the $T_{c,onset}$ [Fig. S5]. Upon cooling toward approximately 110 K, at which the MR begins to emerge [Fig. 3(a)], a pronounced deviation from linearity develops. This nonlinearity in $\rho_{xy}$, together with violations of the Kohler's rule [Fig. S6], indicate the increasing importance of multicarrier transport in the $Fe_i$-modulated BFCA films. The low-field data are then quantitatively analyzed using a semiclassical two-band model [Supplemental Material, Sec. 1, Fig. S7 [51]], from which the carrier densities ($n_e$ and $n_p$) and average mobilities ($\mu_e$ and $\mu_p$) of electrons and holes are extracted and summarized in Fig. 4(b). Below 110 K, electron and hole carriers coexist with comparable densities and exhibit rapidly enhanced mobilities, exceeding 3000 $cm^2$/V·s at 27 K, while their carrier densities remain only weakly temperature dependent. The coincidence between the emergence of these high-mobility charge carriers and the onset of the large linear MR indicates a close correlation between them, while the enhanced carrier mobilities are fully consistent with the ultralow residual resistivity and exceptionally large RRR observed in these films.

Large linear MR has been frequently reported in Dirac and Weyl semimetals [21, 52-54]. In $BaFe_2As_2$-related iron pnictides [17, 18], linear MR was previously attributed to the emergence of two-dimensional Dirac states arising from band folding in the antiferromagnetic phase [55], within the framework of the Abrikosov quantum model [56]. In this scenario, charge carriers undergo cyclotron motion characterized by the magnetic length $l_H \sim \sqrt{\hbar/e\mu_0 H}$ and occupy only the zeroth Landau level in the quantum limit. However, this mechanism is highly unlikely to apply to the present BFCA films, since both parent and doped BFCA films devoid of Fe$_i$ exhibit negligibly small MR [Figs. S4 and S5]. Using $\mu_0 H^* \sim 1.5$ T, the magnetic length is estimated to be $l_H \approx 21$ nm, which largely exceeds half of the film thickness of $\sim 6.6$ nm studied here. Consequently, Landau quantization can not be established, rendering the Abrikosov model inapplicable for accounting for the large linear MR observed under in-plane magnetic fields, particularly in the longitudinal configuration (***H***//***I***). Another frequently invoked explanation for linear MR is the classical Parish Littlewood (PL) model [57], in which spatial inhomogeneity distorts current paths and admixes the Hall response into the longitudinal resistivity, thereby producing apparent MR. However, this model predicts linear MR extending into the very low-field regime [58], a weaker in-plane MR compared with the out-of-plane configuration [59, 60], and even a negative MR for ***H***//***I*** [59], all of which are in contradiction with our observations.

The elimination of these two major mechanisms leaves quantum transport from three-dimensional TDSs in BFCA [9, 61] as the most plausible origin of the anomalous linear MR observed here. This assignment is particularly striking since, in IBSs [9, 10, 16, 61], TDS states usually reside above the Fermi level ($E_F$) and thus contribute only marginally to the low-energy transport, matching with the negligible MR observed in Fe$_i$-free samples [Figs. S4 and S5]. The unique emergence of large linear MR in Fe$_i$-modulated BFCA films indicates that interstitial Fe acts as an effective electronic tuning parameter, driving an upward shift of the $E_F$ and thereby activating the contribution of TDS states to charge transport. This scenario is supported by our DFT calculations. By comparing the band structures of BaFe$_2$As$_2$ [Fig. 4(c)] and BaFe$_{2+\delta}$As$_2$ with $\delta = 0.5$ [Fig. 4(d)], as well as that of the Co-doped compounds [Fig. S8], we find that electron doping predominantly induced by Fe$_i$ shifts the TDS and TI bands downward by approximately 0.21 eV and 0.23 eV, respectively. Notably, while the TSSs associated with TI invariably cross $E_F$ [Fig. S9], the TDS bands are driven across $E_F$ only upon Fe$_i$ incorporation, directly corroborating the attribution of the observed linear positive MR to TDS-driven transport. In parallel, Fe$_i$ incorporation helps maintain antiferromagnetic interactions in BFCA (Tab. S1), which is often regarded as a key prerequisite for the emergence of the topological states [55]. This picture is in agreement with our experimental observation that the unusual quantum transport phenomena emerge below 110 K, matching with the onset of antiferromagnetic fluctuations for optimally doped BFCA [20].

The involvement of TDS-driven transport near the $E_F$ provides a natural explanation for the observed ultralow residual resistivity and large RRR at zero field, owing to the strong suppression of backscattering associated with topological protection [1, 21]. Upon application of a magnetic field, this protection can be lifted via time-reversal-symmetry breaking, giving rise to a large linear MR, as reported in various TDS materials [21, 52, 62]. In addition, the coexistence of high-mobility electrons and holes implies topological transport arising from type-II Dirac fermions, which violate Lorentz invariance and host tilted Dirac cones [21]. While a chiral anomaly in type-I Dirac and Weyl semimetals tends to produce negative longitudinal MR [21, 63-65], magneto-transport in the type-II topological semimetals is substantially more complex and can exhibit unconventional linear positive longitudinal MR [66, 67], as observed. The sign and magnitude of this response are highly sensitive to band tilting, magnetic-field orientation and intervalley scattering [68, 69], warranting further experimental investigation.

In summary, we reveal compelling evidence for topological semimetal transport in IBS films enabled by controlled incorporation of Fe$_i$. Rather than acting as conventional disorder detrimental to superconductivity, Fe interstitials profoundly reconstruct the low-energy electronic structure and activate three-dimensional TDS states deep in the superconducting regime. The demonstrated coexistence of TDSs and superconductivity establishes this system as a promising platform for realizing bulk topological superconductivity and associated Majorana bound states on its side surfaces. Moreover, our systematic control of Fe$_i$ offers a natural explanation for the long-standing and widely scattered magneto-transport results reported across this material family, suggesting that uncontrolled variations in Fe$_i$ content constitute a dominant yet previously overlooked parameter. Together, these findings identify interstitial engineering as a powerful tuning knob for stabilizing and manipulating topological phases in correlated IBSs, opening a new avenue for exploring the interplay between topology, strong correlations, and unconventional superconductivity.

**Acknowledgments**

This work is financially supported by grants from the National Key Research and Development Program of China (Grant No. 2022YFA1403100), the Natural Science Foundation of China (Grant No. 12474130, Grant No. 12134008 and Grant No. 12141403), and the Innovation Program for Quantum Science and Technology (2021ZD0302502).

Z. X. Deng, Q. J. Cheng and J. Jiang contributed equally to this work.

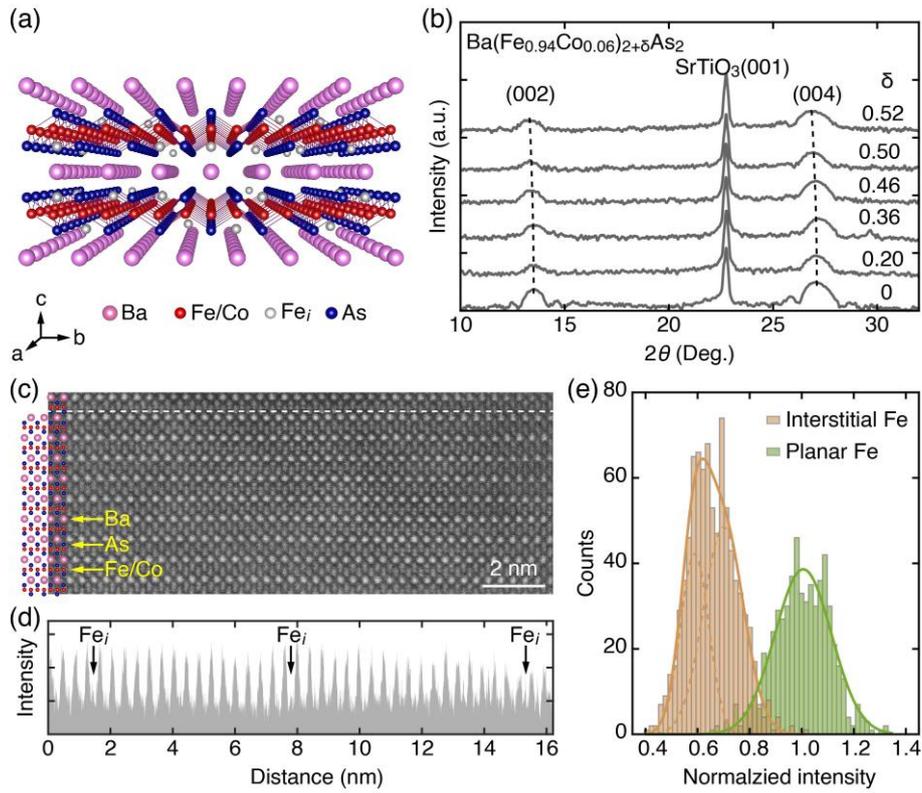

**Figure 1** (a) Schematic crystal structure of BFCA, with randomly distributed Fe$_i$ atoms labeled in gray. (b) XRD spectra of BFCA films with varying nominal Fe$_i$ contents. The dashed lines are guides to the eye. (c) TEM image of an Fe$_i$ ($\delta = 0.52$)-modulated BFCA along the [010] direction, with the corresponding atomic model overlaid on the left. (d) Line profile extracted along the white line in (c), showing site-dependent intensity variations (arrows) due to Fe$_i$ within the As layer. (e) Intensity histograms of Fe$_i$ and planar Fe columns, normalized to the average intensity of the planar Fe columns. The colored lines show the best fit of their intensity distributions to single- and double-Gaussian functions.

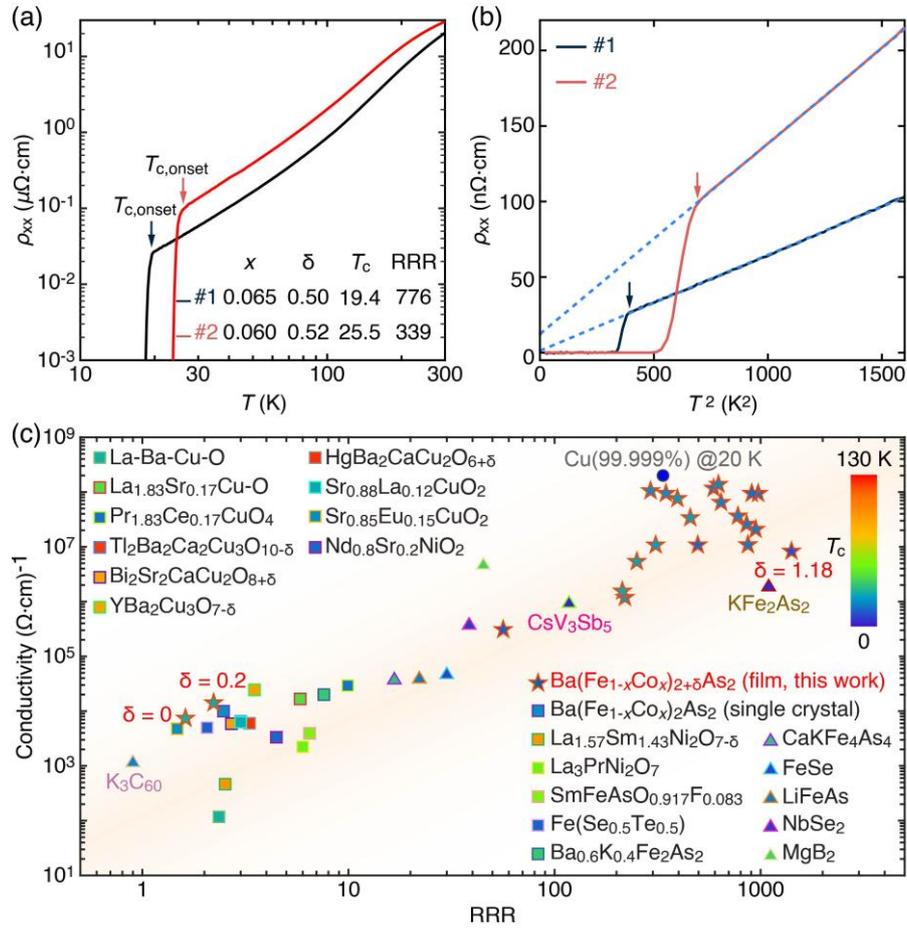

**Figure 2** (a,b) Temperature dependence of the longitudinal resistivity $\rho_{xx}$ for two representative BFCA films (samples #1 and #2) plotted on logarithmic and linear scales, revealing ultralow residual resistivity and large RRR. Blue dashed lines represent linear fits of $\rho_{xx}$ as a function of $T^2$. (c) Electrical conductivity at $T_{c,\text{onset}}$ (i.e. $1/\rho_{xx}(T_{c,\text{onset}})$) *versus* RRR for various superconductors [28-49], with stars marking the BFCA films studied here. The color filling of each symbol encodes the $T_c$ of the corresponding material.

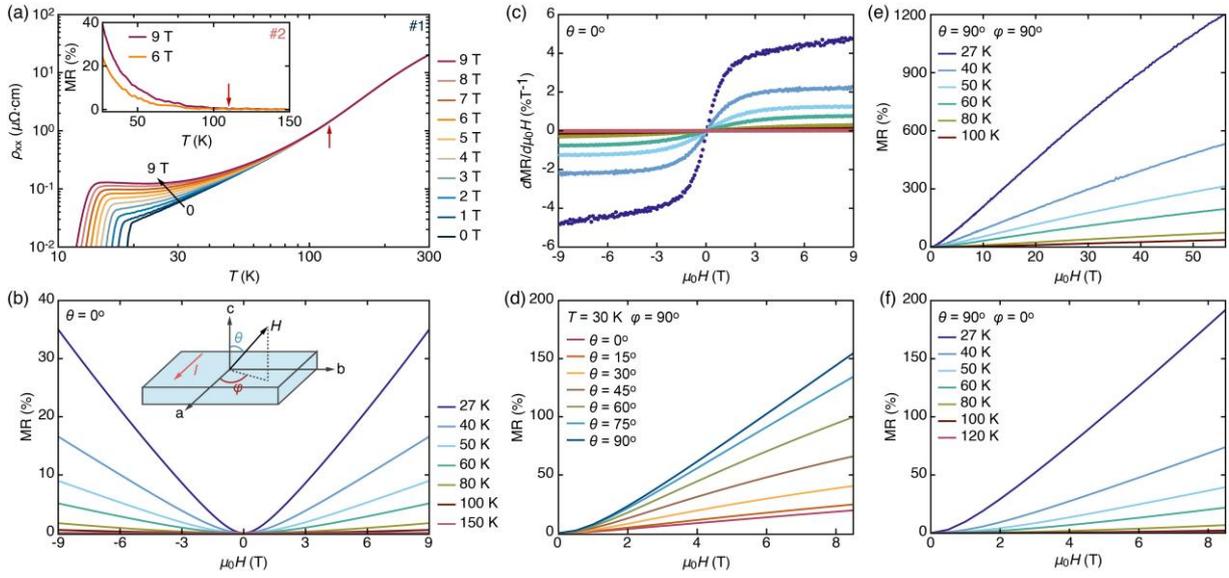

**Figure 3** (a) Temperature-dependent $\rho_{xx}$ of sample #1 measured under perpendicular magnetic fields $\mu_0H$, revealing suppression of superconductivity and the emergence of positive magnetoresistance below 110 K. Inset shows the temperature-dependent MR of sample #2 at two selected magnetic fields. (b,c) Magnetic-field dependence of the transverse MR and its derivative for sample #2 at various temperatures, measured with current ***I*** applied along the *a* axis and the magnetic field along the *c* axis. Inset: schematic of the measurement geometry. (d) Angular dependence of the transverse MR at 30 K, with the magnetic field titled form the *c* toward the *b* axis. (e) Transverse MR measured up to 56 T with the magnetic field applied along the *b* axis. (f) Longitudinal MR measured at various temperatures, with the magnetic field applied along the *a* axis.

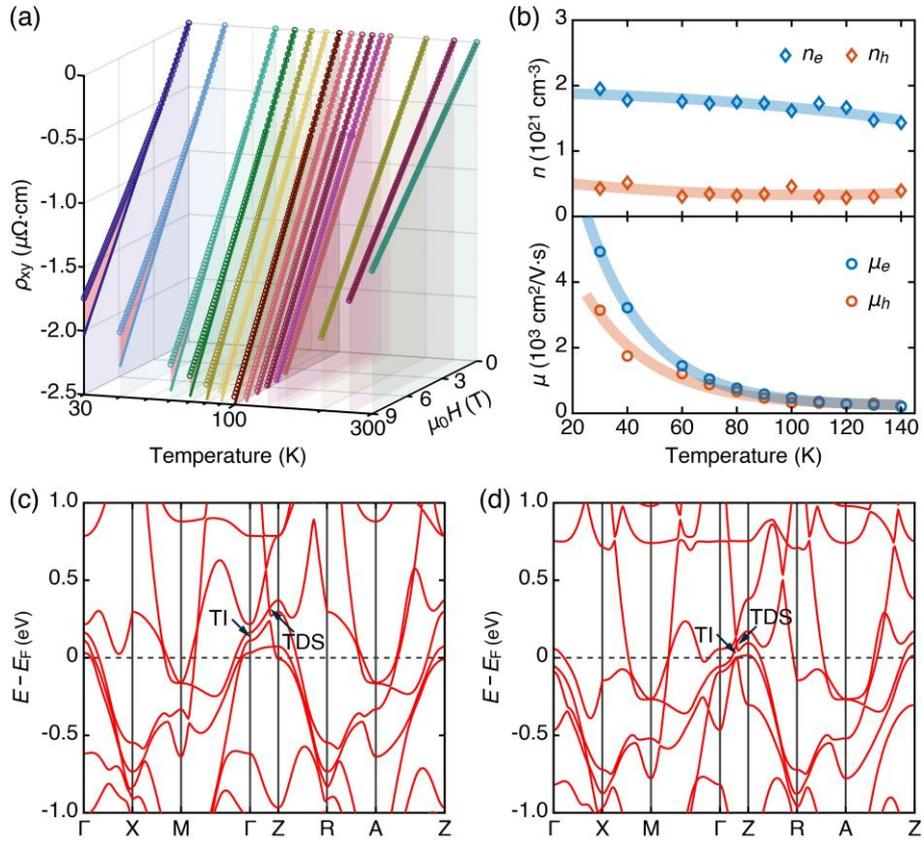

**Figure 4** (a) Temperature dependence of the Hall resistivity $\rho_{xy}$ (circles) in the Fe$_i$-modulated BFCA film ($\delta$ = 0.52), with the solid lines corresponding to linear fits of $\rho_{xy}$ at low magnetic fields. (b) Total carrier densities and average mobilities of electrons and holes extracted from two-band model fits of the low-field $\rho_{xx}$ and $\rho_{xy}$ data. (c,d) Band structures and evolution of topological states in BaFe$_2$As$_2$ and BaFe$_{2+\delta}$As$_2$ ($\delta$ = 0.5).

**Materials and Methods**

**Sample growth.** High-quality $BaFe_2As_2$ and $Ba(Fe_{1-x}Co_x)_{2+\delta}As_2$ epitaxial films were grown by molecular beam epitaxy on undoped, insulating $SrTiO_3(001)$ substrates. All film growth was carried out at a substrate temperature of approximately 500°C under a rich arsenic vapor pressure exceeding $1.0 \times 10^{-7}$ Torr. High-purity Ba (99.99%), Fe (99.9999%), and Co (99.9999%) were co-evaporated from Knudsen effusion cells, with the elemental fluxes calibrated using a quartz crystal microbalance prior to the growth. The Co doping level $x$ (5.0% − 8.1%) and the film thickness were determined from the calibrated Co-to-Fe flux ratio and the deposition time. Before deposition, the $SrTiO_3(001)$ substrates were annealed at approximately 1200°C to obtain atomically flat, $TiO_2$-terminated surfaces. To introduce Fe interstitials into $Ba(Fe_{1-x}Co_x)_{2+\delta}As_2$ films, the growth conditions for stoichiometric $Ba(Fe_{1-x}Co_x)_2As_2$ films were first established. $Fe_i$-modulated $Ba(Fe_{1-x}Co_x)_{2+\delta}As_2$ films were then prepared by selectively reducing the Ba flux while keeping the Fe and Co fluxes unchanged. Under these conditions, excess Fe and Co atoms occupy interstitial sites within the As layers in proportion to their respective fluxes, while the Co substitution level remains essentially unchanged. Owing to the much lower Co flux relative to Fe, the incorporated interstitials are predominantly Fe atoms.

**Electrical transport measurements.** Electrical transport measurements were performed using a standard four-terminal configuration with an excitation current of $I = 1$ $\mu A$ applied along the crystallographic $a$ axis in a Physical Property Measurement System (PPMS, Quantum Design). Electrical contacts were made using silver epoxy. Magnetic fields with magnitudes ranging from $\mu_0 H = 0$ to $\mu_0 H = \pm 9$ T were applied either parallel to the out-of-plane $c$ axis or rotated toward the $ab$ plane, enabling magnetoresistance measurements under both in-plane and out-of-plane field orientations. High-field resistivity measurements were further carried out at the Wuhan National High Magnetic Field Center using pulsed magnetic field up to $\mu_0 H = \pm 56$ T. The longitudinal ($\rho_{xx}$) and transverse Hall ($\rho_{xy}$) resistivity components were obtained by symmetrizing and antisymmetrizing the raw transport data with respect to the magnetic field $\mu_0 H$ according to

$$\rho_{xx} = \frac{\rho_{xx}^+ + \rho_{xx}^-}{2}, \quad \rho_{xy} = \frac{\rho_{xy}^+ - \rho_{xy}^-}{2}, \quad (1)$$

where $\rho_{xx(y)}^+$ and $\rho_{xx(y)}^-$ correspond to the resistivities measured under positive and negative magnetic fields, respectively.

**DFT calculations.** The electronic structures and magnetic properties of $BaFe_2As_2$, with and without $Fe_i$ incorporation, were studied using the DFT calculations within the projector augmented-wave (PAW) method [1], as implemented in the VASP package [2]. The exchange-correlation functional was treated within the generalized gradient approximation (GGA) of the Perdew-Burke-Ernzerhof (PBE) type [3]. The energy cutoff of the plane-wave basis was set to 520 eV, and a 11×11×11 $k$-point mesh was chosen to sample the Brillouin zone (BZ) of primitive cell. Gaussian smearing with a width of 0.05 eV was used for Fermi-surface broadening. The convergence tolerances for the energy and force were set to $10^{-7}$ eV and 0.01 eV/Å, respectively. Topological surface states derived from the topological Dirac semimetal (TDS) and topological insulator (TI) phases of BaFe$_2$As$_2$ and BaFe$_{2+\delta}$As$_2$ were obtained using Wannier90 [4] and Wanniertools [5] packages. The effect of Fe$_i$ at $\delta = 0.5$ was incorporated in the band-structure calculations by introducing electron doping via adjustment of the total electron number in the primitive cell in a compensated jellium background. Magnetic interactions in BaFe$_2$As$_2$ and BaFe$_{2.5}$As$_2$ were examined by comparing the total energies of the stripe antiferromagnetic, Néel antiferromagnetic, and nonmagnetic states, for which a supercell was constructed and an interstitial Fe atom was explicitly introduced to evaluate the energy differences.

**Section 1. Carrier densities and mobilities**

To quantitatively elucidate the charge-carrier dynamics underlying the magneto-transport behavior, we employed a semiclassical two-carrier (electron and hole) model to analyze the measured resistivities of the Fe$_i$-modulated BFCA films. The analysis is primarily restricted to the low-field regime ($\mu_0 H < 1.5$ T), where the magnetoresistance MR exhibits a quadratic field dependence [Fig. 3] and a semiclassical description is applicable. Within this semiclassical two-carrier framework, the longitudinal resistivity $\rho_{xx}(\mu_0 H)$, the zero-field resistivity $\rho_{xx}(0)$, and the Hall resistivity $\rho_{xy}(\mu_0 H)$ are given by

$$\rho_{xx}(\mu_0 H) = \frac{(n_e \mu_e + n_p \mu_p) + (n_e \mu_p + n_p \mu_e)\mu_e \mu_p (\mu_0 H)^2}{e[(n_e \mu_e + n_p \mu_p)^2 + (n_p - n_e)^2 \mu_e^2 \mu_p^2 (\mu_0 H)^2]}, \quad (2)$$

$$\rho_{xx}(0) = \frac{1}{e(n_e \mu_e + n_p \mu_p)}, \quad (3)$$

$$\rho_{xy}(\mu_0 H) = \frac{\mu_0 H [(n_p \mu_e^2 - n_e \mu_p^2) + (n_p - n_e)\mu_e^2 \mu_p^2 (\mu_0 H)^2]}{e[(n_e \mu_e + n_p \mu_p)^2 + (n_p - n_e)^2 \mu_e^2 \mu_p^2 (\mu_0 H)^2]}, \quad (4)$$

where $n_e$ and $n_p$ denote the total electron and hole carrier densities, $\mu_e$ and $\mu_p$ are their respective mobilities, and $e$ is the elementary charge. Notably, the total carrier densities include contributions from all bands in the Fe$_i$-modulated BFCA films, including Dirac-cone-derived states and parabolic bands. Consequently, the mobilities extracted from this analysis should be regarded as effective averaged values.

Based on the definition of the magnetoresistance, MR = [$\rho_{xx}(\mu_0 H) - \rho_{xx}(0)$]/$\rho_{xx}(0)$ can be written as

$$\text{MR} = \frac{n_e n_p \mu_e \mu_p (\mu_e + \mu_p)^2 (\mu_0 H)^2}{(n_e \mu_e + n_p \mu_p)^2}, \tag{5}$$

which captures the experimentally observed quadratic field dependence of the magnetoresistance MR in the low-field limit [Fig. 3]. Accordinlgy, the experimental low-field transport data, including the Hall resistivity $\rho_{xy}(\mu_0 H)$, the zero-field resistivity $\rho_{xx}(0)$, and the MR($\mu_0 H$), were simultaneously fitted using the above expressions, as demonstrated in Fig. S7. From these fits, the carrier densities and mobilities ($n_e$, $n_p$, $\mu_e$, $\mu_p$) were extracted various temperatures. To improve the accuracy and robustness of this multi-parameter fitting, a genetic algorithm was used as a global optimization method [6]. For each dataset, the fitting procedure was repeated multiple times with different initial inputs, and the final fitting parameters were obtained by averaging over the results of these independent runs, thereby reducing statistical fluctuations associated with random initialization.

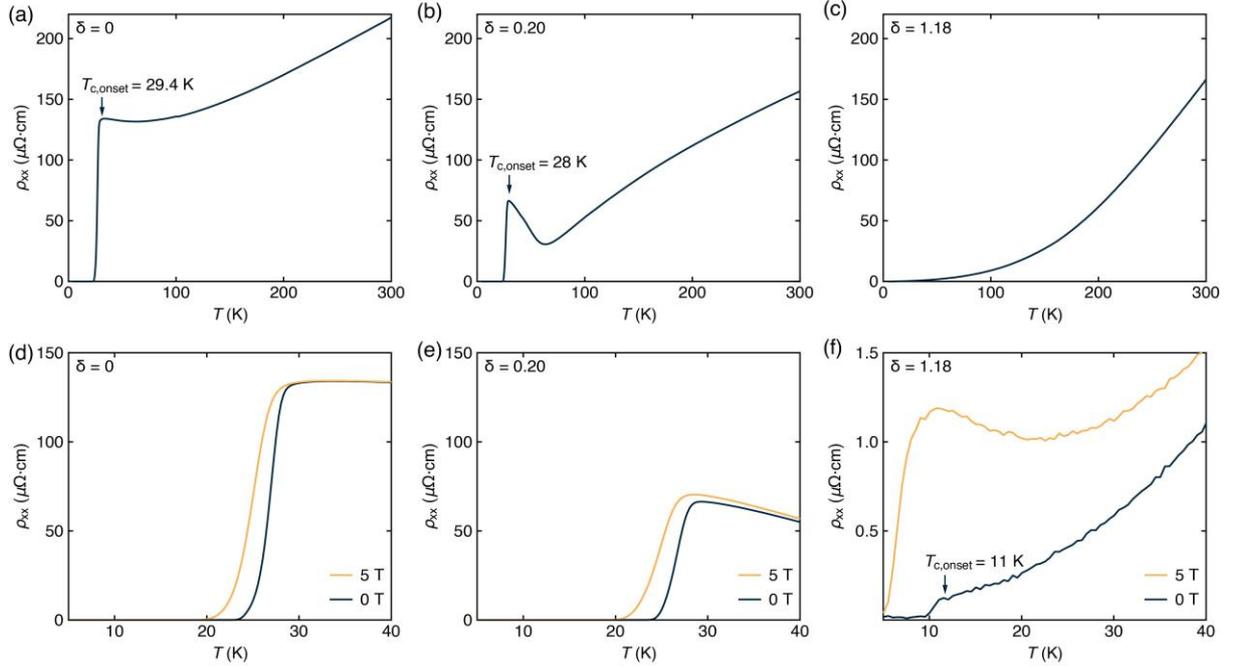

**Fig. S1.** (a-c) Temperature dependence of the longitudinal resistivity $\rho_{xx}$ for three BFCA films with nominal Fe$_i$ contents $\delta = 0$, 0.20, and 1.18, respectively. (d-f) Enlarged views of $\rho_{xx}(T)$ below 40 K, measured at zero magnetic field (black curves) and under a perpendicular magnetic field of 5 T (orange curves). Arrows mark the onset superconducting transition temperature $T_{c,\text{onset}}$, which decreases with increasing $\delta$. A pronounced magnetoresistance is observed in Fe$_i$-modulated BFCA films, whereas no discernible magnetoresistance is detected above $T_{c,\text{onset}}$ for the BFCA film with $\delta \approx 0$.

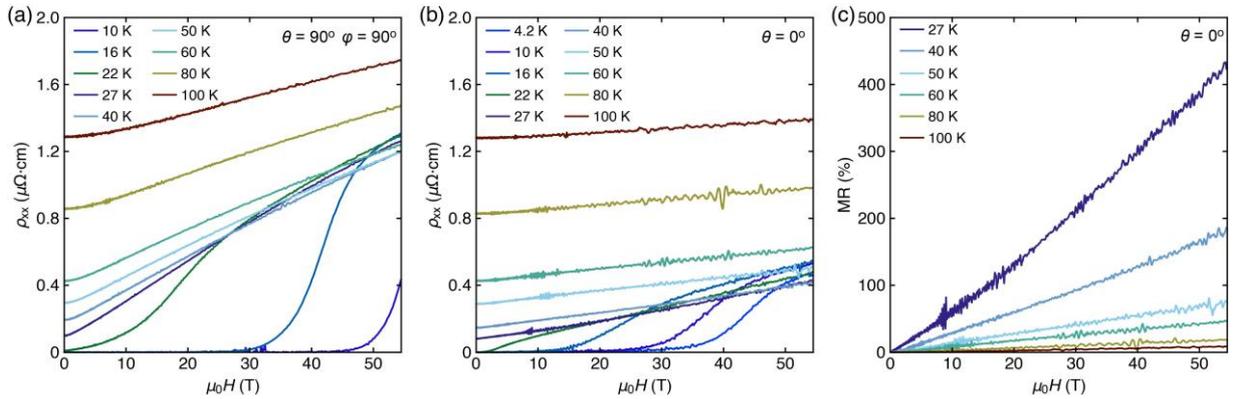

**Fig. S2.** (a,b) Pulsed-field dependence of $\rho_{xx}(\mu_0 H)$ for sample #2, measured at various temperatures with the magnetic field applied along the in-plane $b$ axis and the out-of-plane $c$ axis, respectively. A positive, linear magnetoresistance is observed for both field orientations and is more pronounced when the magnetic field is applied in plane. (c) MR calculated from the data in (b), shown up to 56 T.

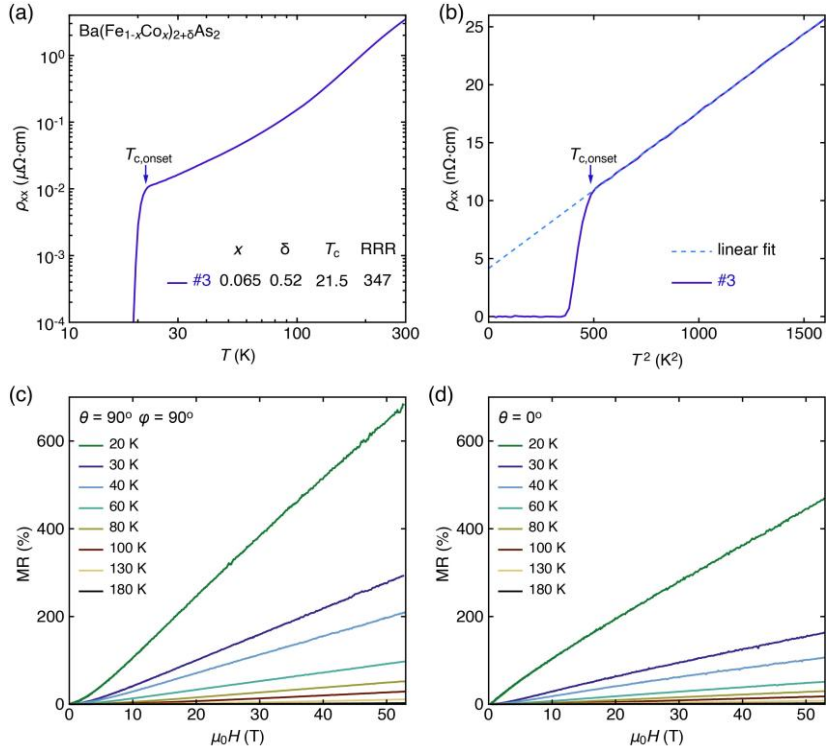

**Fig. S3.** (a,b) Temperature dependence of $\rho_{xx}(T)$ for samples #3 on logarithmic and linear scales, exhibiting an ultralow resistivity of ~ 9.9 n$\Omega$·cm at 21.5 K and a large RRR = 347. The blue dashed line denotes a $T^2$ dependence of $\rho_{xx}$ at low temperatures. (c,d) Pulsed-field dependence of MR($\mu_0H$) for sample #3. A positive, linear MR is observed and is more pronounced when the magnetic field is applied in plane.

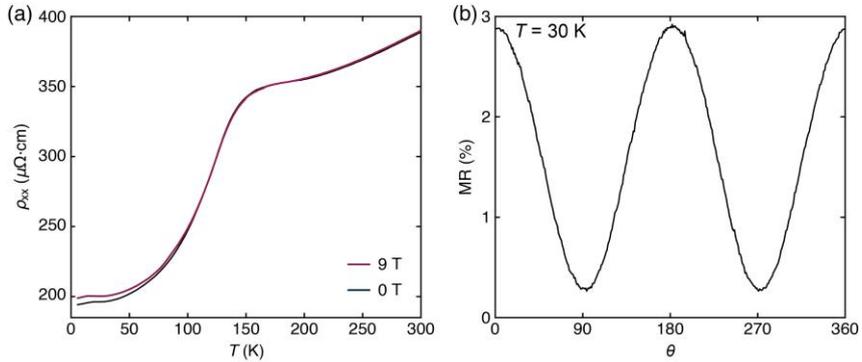

**Fig. S4.** (a) Temperature dependence of $\rho_{xx}(T)$ for stoichiometric BaFe$_2$As$_2$, measured at 0 T (black curve) and 9 T (red curve) with the magnetic field $\mu_0H$ applied along the $c$ axis. The abrupt drop in $\rho_{xx}(T)$ arises from the antiferromagnetic phase transition near 150 K, below which a small magnetoresistance is observed. (b) Polar-angle ($\theta$) dependence of the MR at 30 K, where $\theta = 0°$ corresponds to $\mu_0H \mathbin{/\mkern-6mu/} c$. In contrast to Fe$_i$-

modulated BFCA, the MR in stoichiometric $BaFe_2As_2$ films is significantly smaller and is further reduced by one order of magnitude as the magnetic field is titled from the *c* axis (~ 2.9%) toward the *b* axis (0.3%).

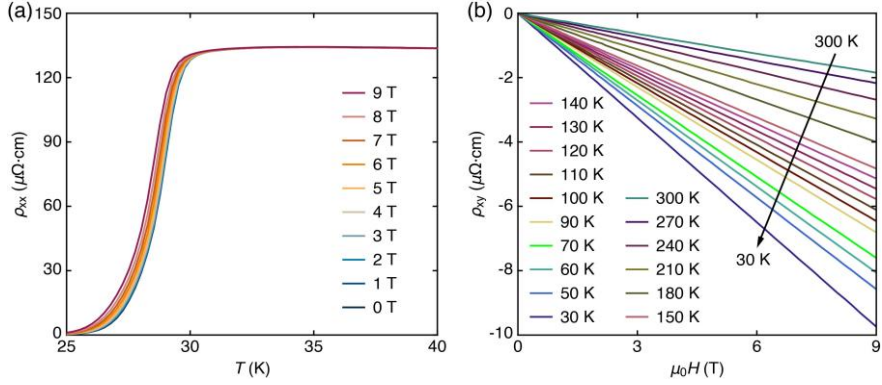

**Fig. S5.** (a) Enlarged views of $\rho_{xx}(T)$ around the onset superconducting transition temperature, measured in the Fe$_i$-free BFCA film ($\delta \approx 0$) under different magnetic fields. (b) Magnetic field dependence of the Hall resistivity $\rho_{xy}(\mu_0H)$ measured at temperatures ranging from 30 K to 300 K in the Fe$_i$-free BFCA film. The Hall resistivity $\rho_{xy}$ scales linearly with $\mu_0H$ over the entire measured field range up to 9 T, with the slope monotonically increasing upon decreasing temperature (see black arrow). The negative sign of $\rho_{xy}$ indicates dominant electron-type charge carriers, consistent with Co substitution for Fe.

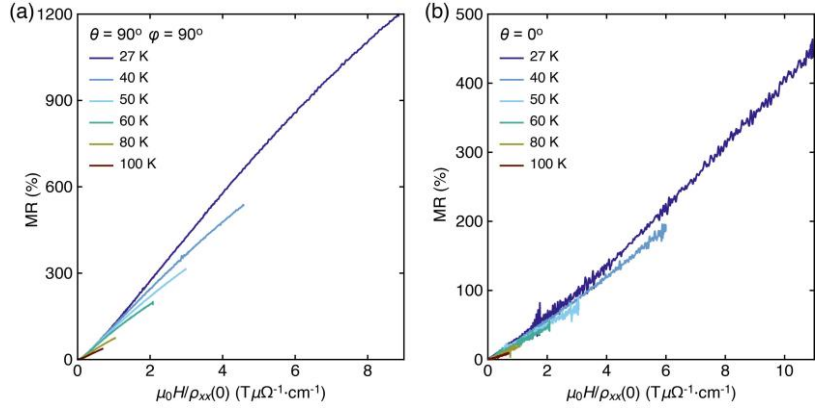

**Fig. S6.** (a) Scaling of the magnetoresistance according to Kohler's rule for sample #2, measured in pulsed magnetic fields up to 56 T applied along the crystallographic *b* and *c* axes, respectively. The Kohler plot exhibits MR = $[\rho_{xx}(\mu_0H) - \rho_{xx}(0)]/\rho_{xx}(0)$ as a function of $\mu_0H/\rho_{xx}(0)$, where $\rho_{xx}(0)$ is the zero-field resistivity. Kohler's rule, which is expected for a single type of charge carrier with a uniform scattering rate across the whole Fermi surface, is violated for both field configurations, as the MR measured at different temperatures fail to collapse onto a single curve. This behavior indicates the presence of multiple charge carriers and/or anisotropic scattering in the BFCA films.

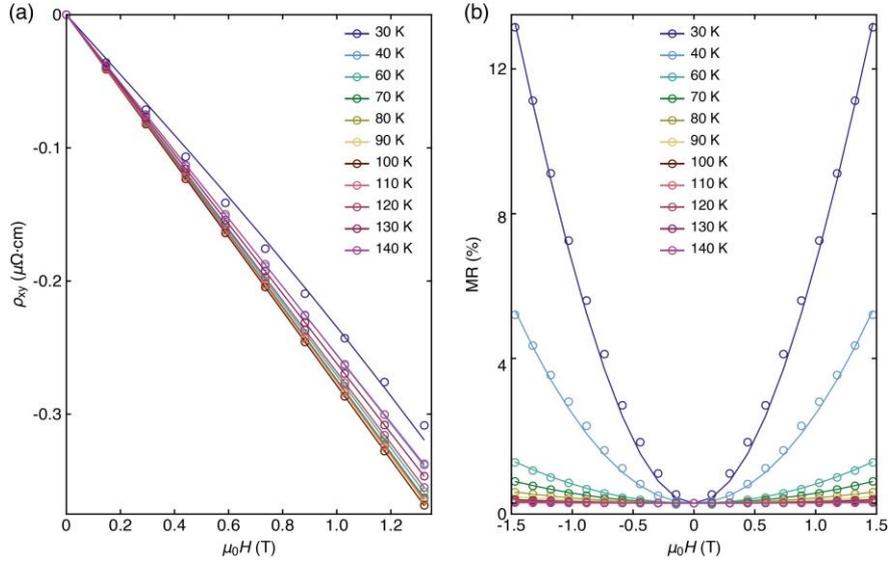

**Fig. S7.** (a,b) Temperature dependence (30 K – 140 K) of (a) the Hall resistivity $\rho_{xy}(\mu_0 H)$ and (b) MR($\mu_0 H$) in the low-field limit ($\mu_0 H < 1.5$ T), measured in the BFCA film with $\delta = 0.52$. The solid lines represent simultaneous fits to $\rho_{xy}(\mu_0 H)$, $\rho_{xx}(0)$ and MR within the semiclassical two-band model.

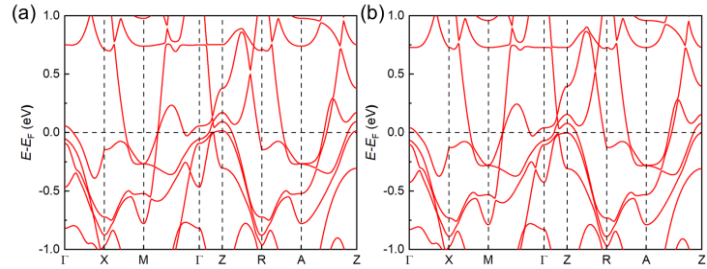

**Fig. S8.** Band structures of (a) $BaFe_{2.5}As_2$ and (b) $Ba(Fe, Co)_{2.5}As_2$ in the nonmagnetic state, where the Co atoms are introduced using the virtual crystal approximation (VCA) method. Due to the very low Co concentration, the two band structures are quite similar.

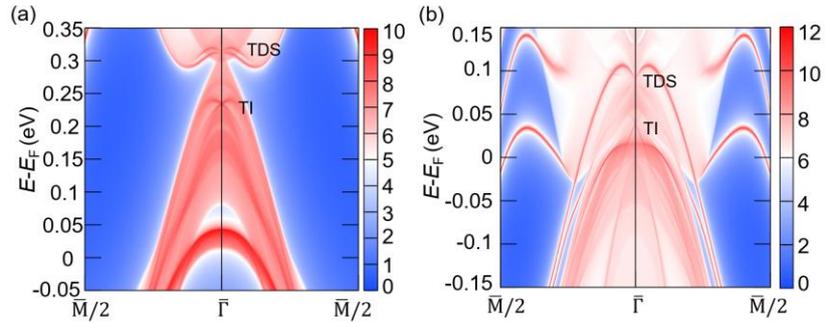

**Fig. S9.** (a,b) Calculated (001) surface spectra of $BaFe_2As_2$ (a) without $Fe_i$ incorporation and (b) with $Fe_i$ incorporation ($\delta = 0.5$), illustrating the evolution of Dirac cones associated with the TDS and TI states.

**Tab. S1.** Comparison of the total energies (in unit of meV/Fe) of the stripe antiferromagnetic state, Néel antiferromagnetic state, and nonmagnetic (NM) state for $BaFe_2As_2$ and $BaFe_{2+\delta}As_2$ with $\delta = 0.5$. The additional $Fe_i$ atoms preserve the stripe-type antiferromagnetic interaction. Notably, a supercell was here constructed and an interstitial Fe atom was explicitly introduced to evaluate the energy differences among different magnetic configurations.

|        | $BaFe_2As_2$ | $BaFe_{2.5}As_2$ |
|--------|--------------|------------------|
| NM     | 60.8         | 242.9            |
| Néel   | 52.1         | 54.9             |
| Stripe | 0.0          | 0.0              |